\documentclass[letterpaper,english,prl,twocolumn,floatfix,footinbib,superscriptaddress,showpacs, showkeys,notitlepage]{revtex4-1}
\usepackage[T1]{fontenc}
\usepackage[latin9]{inputenc}
\usepackage{amsmath}
\usepackage{amssymb}
\usepackage{graphicx}
\usepackage{babel}
\usepackage[colorlinks]{hyperref}
\usepackage[explicit]{titlesec}



\begin{document}

\title{Integrable quantum dynamics of open collective spin models}

\author{Pedro Ribeiro}
\email{ribeiro.pedro@gmail.com}

\selectlanguage{english}%

\affiliation{CeFEMA, Instituto Superior Técnico, Universidade de Lisboa Av. Rovisco
Pais, 1049-001 Lisboa, Portugal}

\affiliation{Beijing Computational Science Research Center, Beijing 100193, China}

\author{Tomaž Prosen}
\email{tomaz.prosen@fmf.uni-lj.si}

\selectlanguage{english}%

\affiliation{Department of Physics, Faculty of Mathematics and Physics, University
of Ljubljana, Ljubljana, Slovenia}
\begin{abstract}
We consider a collective quantum spin-$s$ in contact with Markovian spin-polarized
baths. Using a conserved super-operator charge, a differential
representation of the Liouvillian is constructed to find its exact spectrum
and eigen-modes. We study the spectral properties of the model in
the large-$s$ limit using a semi-classical quantization condition
and show that the spectral density may diverge along certain curves in the complex plane.
We exploit our exact solution to characterize steady-state properties,
in particular at the discontinuous phase transition that arises for
unpolarized environments, and to determine the decay rates of coherences
and populations. Our approach provides a systematic way of finding
integrable Liouvillian operators with non-trivial steady-states as
well as a way to study their spectral properties and eigen-modes.
\noindent\begin{minipage}[t]{1\columnwidth}%
\global\long\def\ket#1{\left| #1\right\rangle }

\global\long\def\bra#1{\left\langle #1 \right|}

\global\long\def\kket#1{\ket{\ket{#1}}}

\global\long\def\bbra#1{\bra{\bra{#1}}}

\global\long\def\braket#1#2{\left\langle #1\right. \left| #2 \right\rangle }

\global\long\def\bbrakket#1#2{\left\langle \left\langle #1\right.\left\Vert #2\right\rangle \right\rangle }

\global\long\def\av#1{\left\langle #1 \right\rangle }

\global\long\def\tr{\text{tr}}

\global\long\def\Tr{\text{Tr}}

\global\long\def\pd{\partial}

\global\long\def\im{\text{Im}}

\global\long\def\re{\text{Re}}

\global\long\def\sgn{\text{sgn}}

\global\long\def\Det{\text{Det}}

\global\long\def\abs#1{\left|#1\right|}

\global\long\def\up{\uparrow}

\global\long\def\down{\downarrow}

\global\long\def\k{\mathbf{k}}

\global\long\def\wks{\mathbf{\omega k}\sigma}

\global\long\def\vc#1{\mathbf{#1}}

\global\long\def\bs#1{\boldsymbol{#1}}

\global\long\def\t#1{\text{#1}}

\global\long\def\L{\mathcal{L}}

\global\long\def\r{\mathcal{\rho}}

\global\long\def\sz{S_{z}}

\global\long\def\sx{S_{x}}

\global\long\def\sy{S_{y}}

\global\long\def\sp{S_{+}}

\global\long\def\sm{S_{-}}

\global\long\def\d{\dagger}

\global\long\def\si{s\rightarrow\infty}
\end{minipage}
\end{abstract}
\maketitle
Understanding the non-equilibrium dynamics of a quantum system coupled
to its environments is of central importance for the possible improvement of current technologies
such as nuclear magnetic resonance, electronic and optical spectroscopy
and inelastic neutron scattering. It is also a key ingredient for a coherent manipulation of quantum states, for classical and quantum
information processing, sensing and metrology. However, modeling the
open quantum dynamics of interacting systems remains a major theoretical
challenge. 

In most cases, the problem can be reduced to the study of a subsystem
of the full system for which the reduced density matrix evolves under
an effective Liouvillian operator, $\pd_{t}\rho=\mathcal{L\left(\rho\right)}$.
When the coupling to the environment is weak and its memory times
are short, the Liouvillian becomes of the Lindblad form \citep{Breuer2002}:
\begin{align}
\mathcal{L} & =\mathcal{L}_{H}+\sum_{\ell}\mathcal{D}_{W_{\ell}},\label{eq:Lindblad}
\end{align}
where $\mathcal{L}_{H}\left(\rho\right)=-i\left[H,\rho\right]$ corresponds
to the unitary evolution under the Hamiltonian $H$, and $\mathcal{D}_{W_{\ell}}\left(\rho\right)=W_{\ell}\rho W_{\ell}^{\dagger}-\frac{1}{2}W_{\ell}^{\dagger}W_{\ell}\rho-\frac{1}{2}\rho W_{\ell}^{\dagger}W_{\ell}$
to the contribution of each dissipative channel by the action of the
jump operator $W_{\ell}$ . In this form the problem becomes amenable
to a number of standard analytic and numeric methods, such as semi-classical, mean-field, or perturbative approximations, Bethe ansatz, exact diagonalization, tensor network methods, etc. 

Remarkably, a number of exact results for model systems have been
recently constructed \citep{Prosen2008a,Prosen2010b,ProsenReview,Prosen2011a,Popkov2013,Prosen2014,Ilievski2017,Znidaric2010,Eisler2011,Medvedyeva2016,Rowlands2017a,Banchi2017,Temme2012,Torres2014}.
For interacting models, there are two known routes to systematically
obtain exact solutions: (i) mapping the Liouvillian to a non-Hermitian
Hamiltonian that acts on two copies of a system for which an exact
solution of the joint problem is known \citep{Eisler2011,Medvedyeva2016,Rowlands2017a,Banchi2017};
or (ii) using a matrix-product operator ansatz to identify the algebraic
structure of the steady-state \citep{ProsenReview,Prosen2011a,Popkov2013,Prosen2014,Ilievski2017}.
All known examples of (i) use Hermitian jump operators which leads
to a maximally-mixed and featureless steady-state but allows to study
the spectrum that determines the dynamics. On the other hand, method
(ii) only provides a solution for the steady-state. The spectrum
and structure of the other eigen-modes remain an open problem. Exact
solutions for both the spectrum and the eigen-modes, that support
a non-trivial steady-state, are only known for quadratic bosonic or 
fermionic models \citep{Prosen2008a,Prosen2010b}.

In this letter, we consider the dissipative dynamics of a quantum spin-$s$, for large $s$, under
a local field $h$ 

\begin{align}
H & =-hS_{z}
\end{align}
 in contact with Markovian spin-polarized baths characterized by the
following jump operators

\begin{align}
W_{0}=\sqrt{\Gamma_{0}}S_{z}; & \quad W_{\pm}=\sqrt{\Gamma\frac{\left(1\mp p\right)}{2}}S_{\pm}.
\end{align}
where $\Gamma_{0}$ characterizes the decoherence rate and $\Gamma\frac{\left(1\mp p\right)}{2}$
are the spin-injection and subtraction rates in a reservoir with a
net polarization $p$. We show that this model admits an exact solution
for the full Liouvillian spectrum and eigen-modes while, at the same time, it
supports a non-trivial steady-state. This solution allows us to determine
the spectral density in the large $s$ limit and to characterize the
steady-state, which undergoes a phase transition where the magnetization
changes discontinuously. This method is particularly useful to compute
the decay rates of coherences and populations at the phase transition
point where perturbative $1/s$ expansions fail. 

A collective spin is described by a single conjugate pair of variables.
Therefore, quantum Hamiltonians of a single spin-$s$ are integrable
\citep{Turbiner1988,Ribeiro2007,Ribeiro2008,Ribeiro2009}. Super-operators,
needed to describe dissipative dynamics, act on a space that is isomorphic
to two copies of the initial Hilbert space. This effective two-variable
problem seems in general not to be integrable. The construction of
our exact solution crucially uses the fact that $\mathcal{L}$ commutes
with a conserved super-operator $\mathcal{Q}_{z}$, that, in the present
case, is simply given by $\mathcal{Q}_{z}\left(\rho\right)=S_{z}\rho-\rho S_{z}$.
This reduces the eigenvalue problem to that of one effective degree
of freedom and generalizes to cases where $\mathcal{Q}_{z}$ is more
complex. 

Collective spin models with Markovian dissipation have been considered
to describe spontaneous emission of ensembles of two-level atoms \citep{Kilin1978,Drummond1978,Drummond1980,Carmichael1999,Schneider2002,Morrison2008,Kessler2012,Hannukainen2017,Iemini2017,Morrison2008}.
Recently, they were also used to model tunneling spectroscopy of atomic
magnets deposited on metallic surfaces in the large bias regime \citep{Shakirov2016,Ferreira}.
For some remarkable cases \citep{Puri1979,Drummond1980}, the steady-state
density matrix can be exactly constructed. Otherwise, when no such
solution is known, the problem is still amenable to semi-classical
methods \citep{Schneider2002,Morrison2008,Kessler2012,Ferreira}.
These studies showed that collective spin models host a number of
phases with qualitatively different steady-states properties and relaxation
regimes. Nevertheless, to our knowledge, an exact solution of the
eigenvalue problem for the Liouvillian operator was not known to date. 

\begin{figure}
\centering{}\includegraphics[width=0.9\columnwidth]{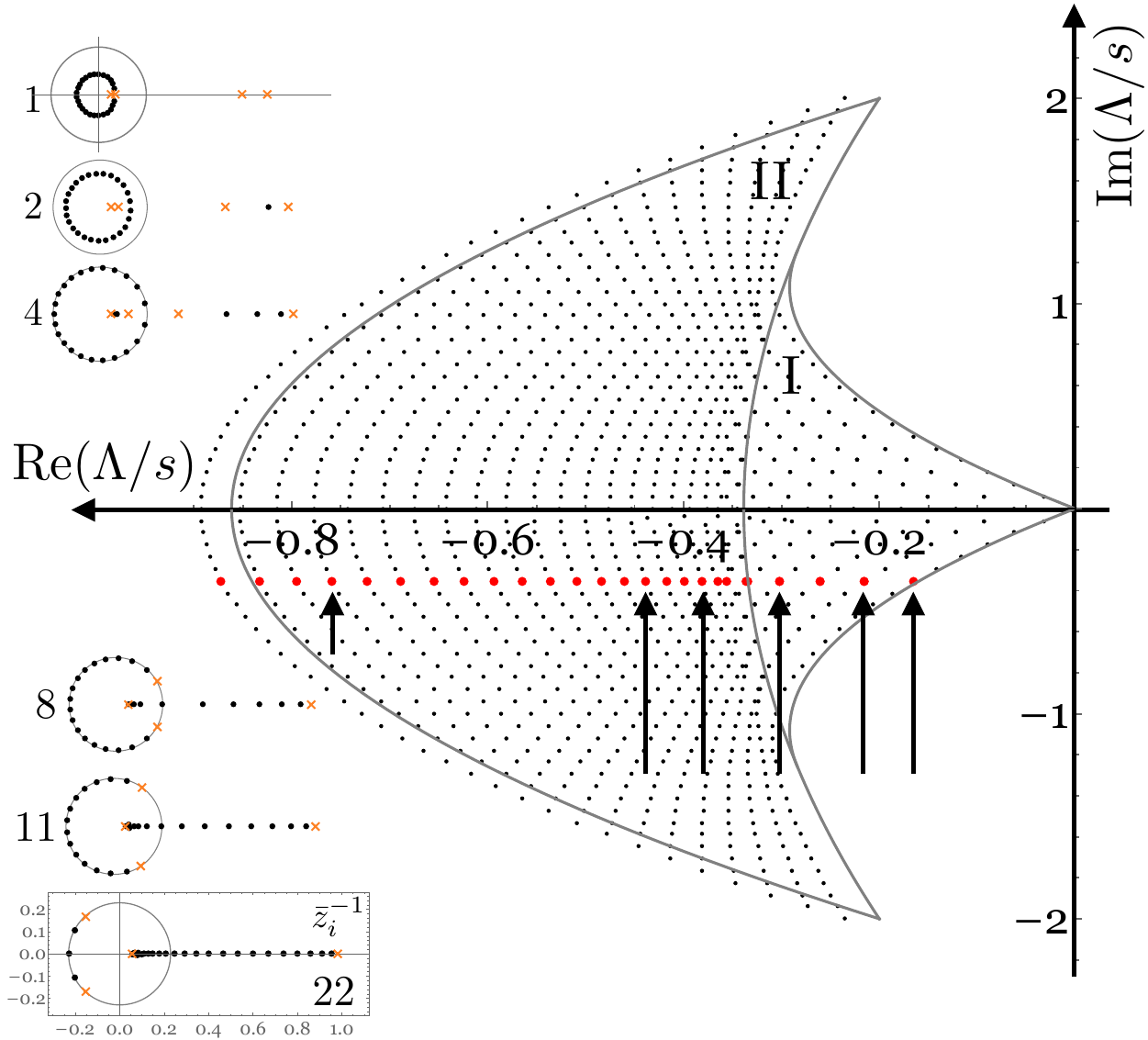}\caption{\label{fig:Spectrum-and-root}Left: Spectrum of the Liouvillian computed
for $h=1$, $\Gamma=1.2$, $p=0.9$, $\Gamma_{0}=0.2$ and $s=17$.
The eigenvalues corresponding to $q=6$ are shown in red. The gray
curves are analytical predictions ($\lambda_k(x ={\rm Im}\Lambda/(2 h s)) = {\rm Re} \Lambda/s$) for the edges of the spectrum for
$s\to\infty$. Insets: Root structure of the eigen-modes. The (inverse)
roots corresponding to some of the eigen-modes are plotted in the
complex plane -- black points. The branch-points $r_{k=1,...,4}$ of
$G_{0}$ are depicted as orange. }
\end{figure}

The Hilbert space of the collective spin is spanned by the Dicke states $\ket{m}$, $m\in\{-s,-s+1,\ldots,s\}$, obeying $S_\pm \ket m = \sqrt{(s\mp m)(s \pm m+1)} \ket{m\pm 1}$, $S_z\ket{m}=m\ket{m}$.
An eigen-mode $\rho$ of $\mathcal{L}$, such that $\mathcal{L}\left(\rho\right)=\Lambda\rho$,
also belongs to an eigen-space of $\mathcal{Q}_{z}\left(\rho\right)=q\rho$
characterized by the eigenvalue $q$ and spanned by the basis $\ket{q+\kappa-s}\bra{\kappa-s}$,
with $\kappa \in\{ 0,1...,2s-q\}$. Thus, $q$ can be seen as a label of a
collection of finite chains and $\kappa$ as the coordinate along
each chain that has a dimension of $2s-q+1$. The operators $S_{-}\rho S_{+}$
and $S_{+}\rho S_{-}$ correspond to nearest neighbor hoppings ($\kappa\to\kappa-1$
and $\kappa\to\kappa+1$) in each finite chain. 

This construction resembles that of an isolated spin-$s$ for which
a representation based on spin coherent states can be used to 
expose the integrable structure of the problem \citep{Ribeiro2007,Ribeiro2008}.
The later relies on the fact that, in spin coherent states basis,
a given state is a polynomial of a single variable on which the Hamiltonian
acts as a differential operator. Here, we proceed in a similar manner,
however not relaying on the $SU(2)$ structure to define the coherent
states. Instead, we define a family of ``coherent''-operators within
the subspace $q$ by 
\begin{align}
\sigma^{\left(q\right)}\left(z\right) & =\sum_{\kappa=0}^{2s-q}c_{q,\kappa}z^{\kappa}\ket{q+\kappa-s}\bra{\kappa-s},
\end{align}
where the coefficients $c_{q,\kappa}=\sqrt{\frac{\left(q+\kappa\right)!\left(2s-\kappa\right)!}{\kappa!\left(2s-q-\kappa\right)!}\frac{\left(2s-q\right)!}{\left(2s\right)!\left(q\right)!}}$
were fixed by requiring that the action of $S_{-}\rho S_{+}$ and
$S_{+}\rho S_{-}$ can be written as differential operators (see
below) and that both operators are extensive in $s$. The inner product
of a generic density matrix with a coherent-operator, $\Psi_{\rho}^{\left(q\right)}\left(\bar{z}\right)=\tr\left\{ \left[\sigma^{\left(q\right)}\left(z\right)\right]^{\dagger}\rho\right\} $,
defines a set of representatives of $\rho$ in the space of polynomials
of $\bar{z}$ of order $2s-q$. In this representation, a super-operator
$\mathcal{O}$ acting on $\rho$ translates to a differential operator
\begin{align}
\Psi_{\mathcal{O}\left(\rho\right)}^{\left(q\right)}\left(\bar{z}\right) & =\mathcal{O}\left(\bar{z},\pd_{\bar{z}}\right)\Psi_{\rho}^{\left(q\right)}\left(\bar{z}\right).
\end{align}
For the diagonal super-operators, i.e. whose action on basis states
are of the form $\mathcal{O}\left(\ket{q+\kappa-s}\bra{\kappa-s}\right)=o\left(\kappa\right)\ket{q+\kappa-s}\bra{\kappa-s}$,
we can simply write $\Psi_{\mathcal{O}\left(\rho\right)}^{\left(q\right)}\left(\bar{z}\right)=o\left(\bar{z}\pd_{\bar{z}}\right)\Psi_{\rho}^{\left(q\right)}\left(\bar{z}\right)$.
The action of the two non-diagonal operators in Eq.\,(\ref{eq:Lindblad}),
i.e. $S_{-}\rho S_{+}$ and $S_{+}\rho S_{-}$, can be obtained by
a straightforward calculation yielding
\begin{align}
\Psi_{S_{-}\rho S_{+}}^{\left(q\right)}\left(\bar{z}\right) & =\pd_{\bar{z}}\left(2s+1-\bar{z}\pd_{\bar{z}}\right)\Psi_{\rho}\left(\bar{z}\right),\\
\Psi_{S_{+}\rho S_{-}}^{\left(q\right)}\left(\bar{z}\right) & =\bar{z}\left(2s-q-\bar{z}\pd_{\bar{z}}\right)\left(\bar{z}\pd_{\bar{z}}+1+q\right)\Psi_{\rho}\left(\bar{z}\right).
\end{align}
In this representation, the eigen-system equation of $\mathcal{L}$
is given by $\mathfrak{L}^{\left(q\right)}\left(\bar{z},\pd_{\bar{z}}\right)\Psi^{\left(q\right)}\left(\bar{z}\right)=\Lambda\Psi^{\left(q\right)}\left(\bar{z}\right)$
with
\begin{align}
\mathfrak{L}^{\left(q\right)}\left(\bar{z},\pd_{\bar{z}}\right) & =sP_{0,0}\left(\bar{z}\right)+P_{0,1}\left(\bar{z}\right)\\
 & +\left[P_{1,0}\left(\bar{z}\right)+\frac{1}{s}P_{1,1}\left(\bar{z}\right)\right]\pd_{z}+\frac{1}{s}P_{2}\left(\bar{z}\right)\pd_{z}^{2}\nonumber 
\end{align}
where\\ 
$P_{0,0}\left(\bar{z}\right)=\frac{q}{2s}\left(-2ih+\Gamma(\frac{q}{2s}-1)((p-1)\bar{z}+1)-\Gamma_{0}\frac{q}{2s}\right)$;
$P_{0,1}\left(\bar{z}\right)=\frac{1}{2}\Gamma(p(\frac{q}{2s}-1)(\bar{z}-1)+(1-\frac{q}{2s})\bar{z}-1)$;
$P_{1,0}\left(\bar{z}\right)=\frac{1}{2}\Gamma\left((p-1)(2\frac{q}{2s}-1)\bar{z}^{2}+p+2(\frac{q}{2s}-1)\bar{z}+1\right)$;
$P_{1,0}\left(\bar{z}\right)=\frac{1}{2}\Gamma(p-1)(\bar{z}-1)\bar{z}$
and $P_{2}\left(\bar{z}\right)=\frac{1}{4}\Gamma\bar{z}(\bar{z}-1)\left[1+p-\left(1-p\right)z\right]$.
Considering the factorizable form of $\Psi^{\left(q\right)}\left(\bar{z}\right)=C\prod_{i=0}^{2s-q}\left(\bar{z}-\bar{z}_{i}\right)$,
with $C$ an non-zero constant, and that the roots $\bar{z}_{i}$
are non-degenerate, we can expand the eigenvalue equation around $\bar{z}_{i}$,
obtaining a set of Bethe-like equations
\begin{align}
\sum_{j\neq i}\frac{1}{\bar{z}_{i}-\bar{z}_{j}} & =\frac{\left(1-p\right)\left(\frac{q}{2}+1\right)}{1+p-\left(1-p\right)\bar{z}_{i}}+\frac{\frac{q}{2}}{1-\bar{z}_{i}}+\frac{s}{\bar{z}_{i}}
\end{align}
whose solution provides the $2s-q$ roots that determine univocally
the eigen-mode of $\mathcal{L}$. Fig.\,\ref{fig:Spectrum-and-root}
shows the spectrum and the root structure for some of the eigen-modes.
The spectrum is constituted of two regions with distinct eigenvalue
distribution. The two regions are separated by a line where eigenvalues
seem to accumulate. The regular spectral structure observed in region
I near $\Lambda=0$ has been identified in Ref. \citep{Ferreira} using
a Holstein-Primakoff transformation to a bosonic system and a subsequent
perturbative expansion in $1/s$. This approach is only able to capture
eigenvalues of order $O\left(s^{0}\right)$ and thus misses the spectral
structure away from the origin. 

The root structure of the eigen-modes, shown as insets in Fig.\,\ref{fig:Spectrum-and-root},
changes depending on which region the corresponding eigenvalues belong
to. In region I, the excitation number of the eigen-mode can be obtained by
counting the number of roots $n_{\t I}$ that lie within the two disconnected
segments on the real axes depicted in blue. In region II, the excitation
number is given by $2s-q-n_{\t{II}}$, where $n_{\t{II}}$ is the
number of roots that lie along a circle around the origin depicted
as a gray line. 

In order to study the spectrum in the large $s$ limit it is useful
to derive a Riccati-like equation for the logarimic derivative of
$\Psi_{\rho}\left(\bar{z}\right)$: $G\left(\bar{z}\right)=\frac{1}{2s-q}\pd_{z}\ln\Psi_{\rho}\left(\bar{z}\right)=\frac{1}{2s-q}\sum_{i}\left(\bar{z}-\bar{z}_{i}\right)^{-1}$.
For this quantity, the contour integral around a closed path $\gamma$
\begin{align}
\int_{\gamma}d\bar{z}\,G\left(\bar{z}\right) & =\frac{2\pi i}{2s-q}n,
\end{align}
 is quantized, with $n$ being the number of zeros of $\Psi_{\rho}\left(\bar{z}\right)$
encircled by $\gamma$. This quantization condition can be used to
fix the real part of $\Lambda$. The imaginary part is fixed by the
sector of $\mathcal{Q}_{z}$: $\t{Im}\left(\Lambda/s\right)=-qh$. 

\begin{figure}
\begin{centering}
\includegraphics[width=0.9\columnwidth]{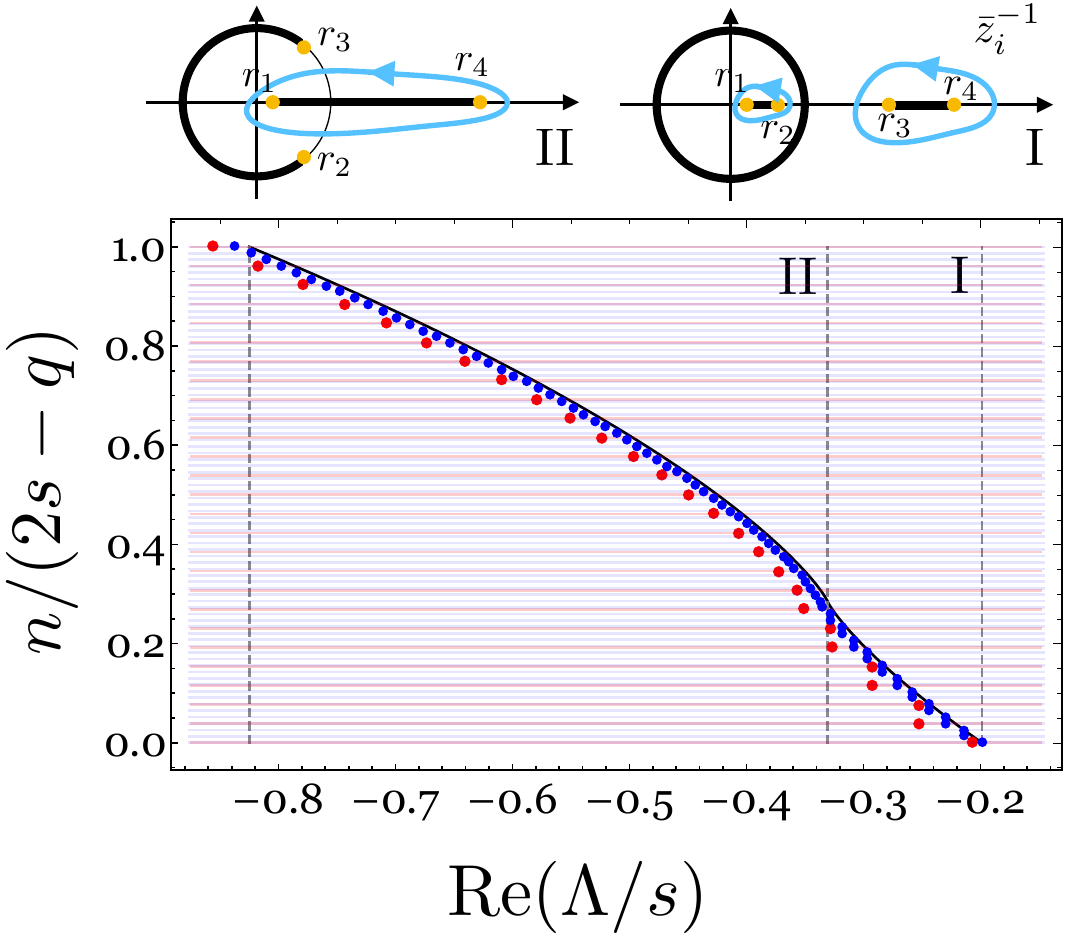}\caption{\label{fig:quantization}Upper panel: contours of integration taken
to obtain the quantization condition in region I and II of the spectrum.
Lower panel: Comparison between the numerics (red dots for $s=17$
and blue dots for $s=50$) and the leading order analytic prediction
(crossings between the horizontal red or blue lines with the black
curve) computed for $h=1$, $\Gamma=1.2$, $p=0.9$, $\Gamma_{0}=0.2$,
and $q=5$. }
\par\end{centering}
\end{figure}
Setting $\Lambda=s\left(\lambda-i2hx\right)$ and $q=2sx$, where
$\lambda\in\mathbb{R}$ and $x\in[0,1]$ are of order zero in $s$,
and expanding $G\left(\bar{z}\right)=G_{0}\left(\bar{z}\right)+\frac{1}{s}G_{1}\left(\bar{z}\right)+...$,
we obtain a set of equations that can be solved hierarchically for
each power of $s$. The leading order term yields an algebraic relation
for $G_{0}\left(\bar{z}\right)$:
\begin{align}
\lambda+i2hx & =P_{0,0}\left(\bar{z}\right)+2\left(1-x\right)P_{1,0}\left(\bar{z}\right)G_{0}\left(\bar{z}\right)\nonumber \\
 & +4\left(1-x\right)^{2}P_{2}\left(\bar{z}\right)G_{0}^{2}\left(\bar{z}\right).
\end{align}
The solution can be put in the form $G_{0}\left(\bar{z}\right)=\left[Q\left(\bar{z}\right)\pm\sqrt{W\left(\bar{z}\right)}\right]/D\left(\bar{z}\right)$,
where $Q\left(\bar{z}\right)$ and $D\left(\bar{z}\right)$ are second
and third-order polynomials in $\bar{z}$ and $W\left(\bar{z}\right)=\prod_{k=1}^{4}\left(\bar{z}-r_{k}^{-1}\right)$
is a fourth order with polynomial roots $r_{k}^{-1}$. The structure
of these roots determines the boundaries of the branch cuts of
$G_{0}\left(\bar{z}\right)$. The cuts can be seen as the results
of the accumulation of the poles $\bar{z}_{i}$ along certain lines
in the complex plane. Accidents in the spectrum, such as spectral
boundaries and lines where the spectrum changes nature arise when
two or more of the roots meet, in which case $\pd_{\bar{z}}W\left(\bar{z}_{i}\right)=0$.
This condition is used to determine the curves seen in Fig.\,\ref{fig:Spectrum-and-root}, which can be parametrized as $\lambda_{k}(x)$.
The roots $r_{k}$ are depicted in orange on the right panel of the
same figure. 

At leading order, the quantization condition, given by $\int_{\gamma}d\bar{z}\,G_{0}\left(\bar{z}\right)=\frac{2\pi i}{2s-q}n$,
fixes the value of $\lambda$ as a function of the number of roots
inside $\gamma$. For the particular example given here we choose
$\gamma$ as in Fig.\,\ref{fig:quantization}(upper panel). In this
way the $n$-th eigen mode has exactly $n$ roots inside the cut.
A comparison between the values of $\lambda$ obtained imposing this
leading order quantization condition and the numerically exact results
obtained by exact diagonalization of the Liouvillian is given in Fig.\,\ref{fig:quantization}(lower panel). 

\begin{figure}
\centering{}\includegraphics[width=0.9\columnwidth]{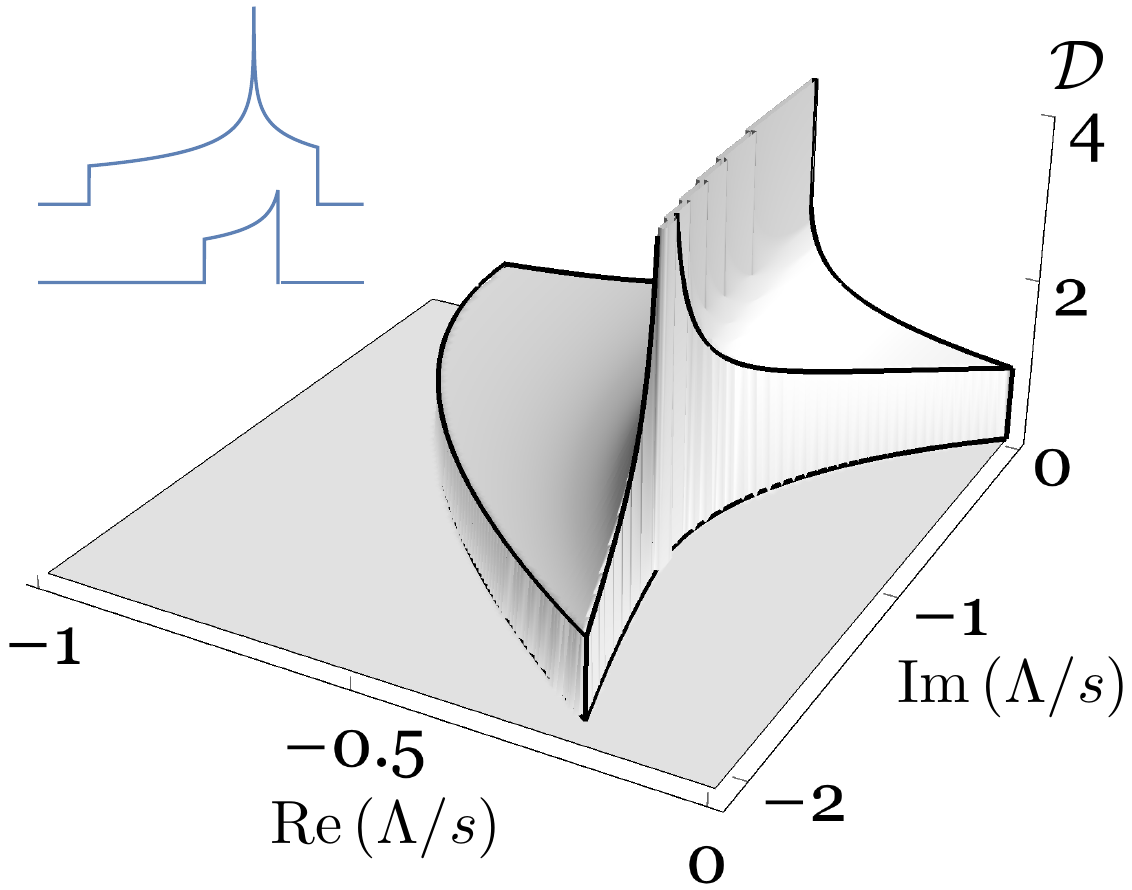}\caption{\label{fig:DOS} Density of eigenvalues of the Liouvillian in the
thermodynamic limit, $\mathcal{D}$, plotted for $h=1$, $\Gamma=1.2$,
$p=0.9$, $\Gamma_{0}=0.2$ and $s=17$. Inset: cut of the 3d plot
for $\Lambda/s=-0.3$ (upper) and $\Lambda/s=-1.5$ (lower). }
\end{figure}

The density of eigen-modes as a function of $\lambda$ defined as
$D_{x}\left(\lambda\right)=\frac{\left(1-x\right)}{2\pi i}\pd_{\lambda}\int_{\gamma}d\bar{z}\,G_{0}\left(\bar{z}\right)$,
normalized such that $\int_{{\rm min}_k \lambda_k(x)}^{{\rm max}_k \lambda_k(x)}d\lambda\,D\left(\lambda\right)=1-x$
is given by particularly simple expressions 
\begin{align}
D_{x}^{\text{I}}\left(\lambda\right)= & \frac{4}{\pi\Gamma(p+1)}\frac{K\left[\frac{\left(r_{1}-r_{2}\right)\left(r_{3}-r_{4}\right)}{\left(r_{3}-r_{2}\right)\left(r_{1}-r_{4}\right)}\right]}{\sqrt{\left(r_{2}-r_{3}\right)\left(r_{1}-r_{4}\right)}}
\end{align}
and 
\begin{align}
D_{x}^{\text{II}}\left(\lambda\right)= & \frac{2}{\pi\Gamma(p+1)}\frac{\tilde{K}\left[\frac{\left(r_{2}-r_{3}\right)\left(r_{1}-r_{4}\right)}{\left(r_{1}-r_{3}\right)\left(r_{2}-r_{4}\right)}\right]}{\sqrt{\left(r_{1}-r_{3}\right)\left(r_{4}-r_{2}\right)}}
\end{align}
in terms of the complete elliptic integral of the first kind $K(z)=\int_{0}^{\frac{\pi}{2}}\frac{1}{\sqrt{1-z\sin^{2}t}}dt$,
and where $\tilde{K}\left(z\right)=K(z)-2iK(1-z)$ is obtained from
$K(z)$ by analytic continuation, changing its branch cut from $\left(1,\infty\right)$
to $\left(-\infty,1\right)$. The density of eigenvalues, $\mathcal{D}\left(\Lambda/s\right)=D_{-\frac{\im\Lambda}{2hs}}\left(\frac{\re\Lambda}{s}\right)$,
in the complex $\Lambda/s$ plane is depicted in Fig\,\ref{fig:DOS}.
The logarithmic divergence of $\mathcal{D}$ along the line separating
regions I and II signals an accumulation of eigenvalues at these points.
The inset, in the upper-left part of Fig.\,\ref{fig:DOS}, shows
two cuts at fixed $\im\Lambda/s$: in the upper case $\mathcal{D}$
has support in I and II and a divergence seen when the separating
line is crossed; in the lower, $\mathcal{D}$ has support only in
II. Note that for the point $p=0$, the region I vanishes. In this special
case (see below) there is a square-root divergence of $\mathcal{D}$
for $\Lambda/s\to0$ instead of the logarithmic accumulation observed
for finite $p$. 

\begin{figure}
\centering{}\includegraphics[width=1\columnwidth]{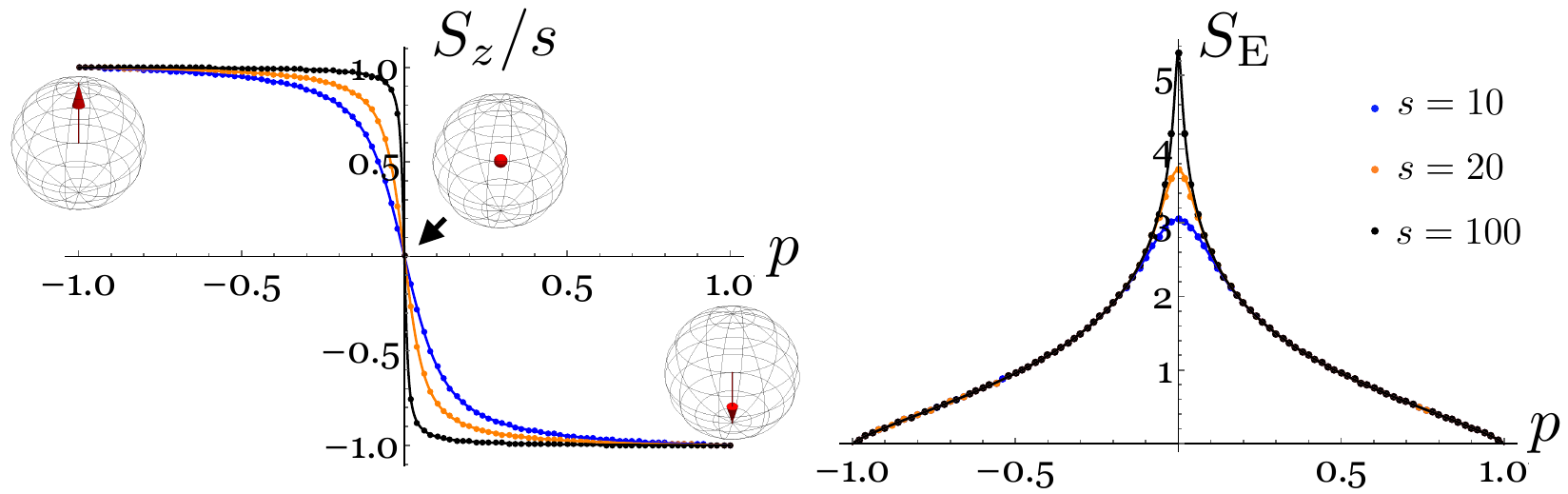}\caption{\label{fig:ss} Steady-state magnetization along the $z$ direction
(left) and entropy (right) of the non-equilibrium steady-state as
a function of $p$ for $h=1$, $\Gamma=1.2$, $p=0.9$, $\Gamma_{0}=0.2$
and several values of $s$. Dots are numerical data obtained by exact
diagonalization and lines are analytic results. }
\end{figure}

Let us now turn to the steady-state properties. The steady-state density
matrix $\rho_{0}$ corresponds to the zero eigenvalue of the Liouvillian
($\Lambda_{0}=0$) and belongs to the sector $q=0$. Within this sector
$c_{q=0,\kappa}=1$, therefore, for a density matrix $\rho=\sum_{\kappa=0}^{2s}w_{\kappa}\ket{\kappa-s}\bra{\kappa-s}$
in this sector, the corresponding polynomial is simply given by $\Psi_{\rho}^{\left(0\right)}\left(\bar{z}\right)=\sum_{\kappa=0}^{2s}w_{\kappa}z^{\kappa}$.
This implies that $\Psi_{\rho}^{\left(0\right)}\left(\bar{z}=1\right)=\tr\left[\rho\right]$.
The steady-state polynomial representation can be obtained by solving
the differential equation $\mathfrak{L}^{\left(0\right)}\left(\bar{z},\pd_{\bar{z}}\right)\Psi_{\rho_{0}}^{\left(0\right)}\left(\bar{z}\right)=0$,
imposing that the solution is a polynomial in $\bar{z}$ normalized
such that $\Psi_{\rho}^{\left(0\right)}\left(1\right)=1$ . In this
way we obtain 
\begin{align}
\Psi_{\rho_{0}}^{\left(0\right)}\left(\bar{z}\right) & =\frac{z_{p}-1}{z_{p}^{2s+1}-1}\frac{\left(\bar{z}z_{p}\right)^{2s+1}-1}{\bar{z}z_{p}-1}
\end{align}
with $z_{p}=\frac{1-p}{1+p}$. It worth noting that this solution
is equivalent to taking $w_{\kappa}\propto z_{p}^{\kappa}$. With the
explicit expression of the steady-state we can now compute its properties.
Fig.\,\ref{fig:ss} (left panel) shows the mean value of 
\begin{align}
\av{S_{z}} & =\Psi_{S_{z}\rho_{0}}^{\left(0\right)}\left(\bar{z}=1\right)=\frac{\left(2s+1\right)}{1-z_{p}^{-\left(2s+1\right)}}-\frac{1}{1-z_{p}^{-1}}-s
\end{align}
 as a function of $p$ together with numerical data obtained by exact
diagonalization of the Liouvilian. At the thermodynamic limit there
is a discontinuous transition in the spin polarization at $p=0$.
This can also be seen in the steady-state entropy, defined as $S_{E}=-\tr\left[\rho_{0}\ln\rho_{0}\right]$,
that can also be simply computed 
\begin{align}
S_{E} & =\log\left(\frac{z_{p}^{2s+1}-1}{z_{p}-1}\right)\nonumber \\
 & +\frac{z_{p}\left\{ \left[1-2s\left(z_{p}-1\right)\right]z_{p}^{2s}-1\right\} \log(z_{p})}{\left(z_{p}-1\right)\left(z_{p}^{2s+1}-1\right)}.
\end{align}
As show in Fig.\,\ref{fig:ss} (cental panel) this quantity has a
maximum for $p=0$ where the steady-state is proportional to the unit
matrix, for which case $S_{E}=\ln\left(2s+1\right)$ is maximal. Away
from $p=0$, the thermodynamic limit value of $S_{E}$ is finite and
vanishes for $p=\pm1$ were the steady-state is pure and corresponds
to that of a fully polarized spin. 

Another important quantity is the spectral gap given by the first
non-zero eigenvalue $\Delta=-\re\left(\min_{n\neq0}\Lambda_{n}\right)$.
$\Delta$ dominates the asymptotic long time decay of the dynamics
to the steady-state. For $p\neq0$ this quantity attains a finite
value in the thermodynamic limit, given by $\lim_{s\to\infty}\Delta=\abs p\Gamma/2$
which can be computed by the method of Ref. \citep{Ferreira}. For
$p=0$, $\Delta$ vanishes in the thermodynamic limit. This is expected
since at this point region I vanishes and the spectrum is of a different
nature. In order to gain some insight to the spectrum at this special
point we look for solutions of the eigenvalue condition $\mathfrak{L}^{\left(0\right)}\left(\bar{z},\pd_{\bar{z}}\right)\Psi_{\rho_{n}}^{\left(0\right)}\left(\bar{z}\right)=\Lambda_{n}^{\left(0\right)}\Psi_{\rho_{n}}^{\left(0\right)}\left(\bar{z}\right)$
at $p=0$ imposing the eigenvector to be a polynomial of order at
most $2s$ in $\bar{z}$. Remarkably, at this special point the eigenvalue
problem can be reduced to the solution of the hypergeometric differential
equation and we have $\Psi_{\rho_{n}}^{\left(0\right)}\left(\bar{z}\right)=(\bar{z}-1)^{n}\,_{2}F_{1}(n+1,n-2s;2(n+1);1-\bar{z})$,
with $_{2}F_{1}$ the hypergeometric function, and 
\begin{align}
\Lambda_{n}^{\left(0\right)} & =-\frac{\Gamma}{4s}n(n+1)
\end{align}
with $n\in\mathbb{N}_{0}$. The same procedure for sectors $q=\pm1$
yields $\Lambda_{n}^{\left(\pm1\right)}=\mp ih-\frac{\Gamma+\Gamma_{0}+\Gamma n(n+3)}{4s}$.
Therefore, we find that the timescales dominating the decay of the populations
and the coherences, respectively $T_{1}=\abs{\re\Lambda_{1}^{\left(0\right)}}^{-1}=\frac{2s}{\Gamma}$
and $T_{2}=\abs{\re\Lambda_{0}^{\left(1\right)}}^{-1}=\frac{4s}{\Gamma+\Gamma_{0}}$,
diverge with $s$. 

In conclusion, we provide an exactly solvable case of a dissipative
system with a non-trivial steady-state where the spectral properties and eigen-modes can be systematically studied in the semiclassical regime of large spin. 
The example provided in this letter explicitly shows that conserved super-operator 
charges can be used to construct exact solutions of integrable Liouvillians
in the same manner conserved quantities do for integrable Hamiltonians.
This construction provides another route for finding exactly solvable
models of dissipative open systems.
\begin{acknowledgments}
PR acknowledges support by FCT through the Investigador FCT contract
IF/00347/2014 and Grant No. UID/CTM/04540/2013. TP acknowledges ERC Advanced grant 694544 -- OMNES.
\end{acknowledgments}

\bibliographystyle{apsrev4-1}
\bibliography{Open_Lipkin}

\end{document}